\documentstyle[12pt]{article}
\def\newpic#1{%
   \def\emline##1##2##3##4##5##6{%
      \put(##1,##2){\special{em:point #1##3}}%
      \put(##4,##5){\special{em:point #1##6}}%
      \special{em:line #1##3,#1##6}}}
\newpic{}
\newcommand{\be}{\begin{equation}}
\newcommand{\ee}{\end{equation}}
\newcommand{\ba}{\begin{eqnarray}}
\newcommand{\ea}{\end{eqnarray}}
\newcommand{\baa}{\begin{eqnarray*}}
\newcommand{\eaa}{\end{eqnarray*}}
\newcommand{\bb}{\begin{thebibliography}}
\newcommand{\eb}{\end{thebibliography}}
\newcommand{\ci}[1]{\cite{#1}}

\newcommand{\lab}[1]{\label{#1}}
\newcommand{\re}[1]{(\ref{#1})}
\newcommand{\nn}{$1/n$}
\newcommand{\tr}{{\rm tr\,{\bf 1}}}
\newcommand{\la}{\langle}
\newcommand{\ra}{\rangle}
\newcommand{\laa}{\langle \! \langle}
\newcommand{\raa}{\rangle \! \rangle}

\newcounter{hran}
\newcommand{\eqalph}{\setcounter{hran}{\value{equation}}
\refstepcounter{hran}\setcounter{equation}{0}
\renewcommand{\theequation}{\thehran\alph{equation}}}

\newcommand{\eqarab}{\setcounter{equation}{\value{hran}}
\renewcommand{\theequation}{\arabic{equation}}}

\newcommand{\al}{\ifmmode\alpha\else$\alpha$\fi}
\newcommand{\psb}{\ifmmode\overline{\psi}\else$\overline{\psi}$\fi}
\newcommand{\pl}{\ifmmode\partial\else$\partial$\fi}
\newcommand{\plh}{\ifmmode\widehat{\pl}\else$\widehat{\pl}$\fi}
\newcommand{\gm}{\ifmmode\gamma\else$\gamma$\fi}
\newcommand{\Gm}{\ifmmode\Gamma\else$\Gamma$\fi}
\newcommand{\eps}{\ifmmode\epsilon\else$\epsilon$\fi}
\newcommand{\bt}{\ifmmode\beta\else$\beta$\fi}
\newcommand{\gps}{\ifmmode\gm_{\psi}\else$\gm_{\psi}$\fi}
\newcommand{\gpsb}{\ifmmode\gm_{\psb}\else$\gm_{\psb}$\fi}
\newcommand{\sg}{\ifmmode\sigma\else$\sigma$\fi}
\newcommand{\dl}{\ifmmode\delta\else$\delta$\fi}
\newcommand{\Dl}{\ifmmode\Delta\else$\Delta$\fi}
\newcommand{\Dps}{\ifmmode\Dl_{\psi}\else$\Dl_{\psi}$\fi}
\newcommand{\dps}{\ifmmode d_{\psi}\else$d_{\psi}$\fi}
\newcommand{\Dpsb}{\ifmmode\Dl_{\psb}\else$\Dl_{\psb}$\fi}
\newcommand{\dpsb}{\ifmmode d_{\psb}\else$d_{\psb}$\fi}
\newcommand{\Dsg}{\ifmmode\Dl_{\sg}\else$\Dl_{\sg}$\fi}
\newcommand{\om}{\ifmmode\omega\else$\omega$\fi}
\newcommand{\et}{\ifmmode\eta\else$\eta$\fi}
\newcommand{\vph}{\ifmmode\varphi\else$\varphi$\fi}
\newcommand{\LL}[1]{\ifmmode{\cal L}_{#1}\else${\cal L}_{#1}$\fi}
\renewcommand{\v}{\ifmmode{\cal V}\else${\cal V}$\fi}

\newcommand{\th}[2]{\ifmmode\theta^{#1}_{#2}\else$\theta^{#1}_{#2}$\fi}
\newcommand{\Ph}[1]{\ifmmode\Phi^{#1}\else$\Phi^{#1}$\fi}
\newcommand{\J}[2]{\ifmmode J^{#1}_{#2}\else$J^{#1}_{#2}$\fi}
\newcommand{\NN}[1]{\ifmmode N^{#1} \else${N}^{#1}$\fi}
\newcommand{\T}[1]{\ifmmode T^{#1}\else$T^{#1}$\fi}
\renewcommand{\S}[1]{\ifmmode S_{#1}(\vph)\else$S_{#1}(\vph)$\fi}
\newcommand{\ld}[1]{\ifmmode\lambda_{#1}\else$lambda_{#1}$\fi}
\newcommand{\DA}[1]{\ifmmode {\cal D}_A^{#1}\else${\cal D}_A^{#1}$\fi}
\newcommand{\OO}[1]{\ifmmode {\cal O}_{#1}\else${\cal O}_{#1}$\fi}
\title{On Calculation of $1/n$ Expansions of Critical Exponents in
the Gross--Neveu Model with the Conformal Technique.}
\author{{ Derkachov S.E.}\\
{\small \it St.-Petersburg Technology Institute, St.-Petersburg, Russian
Federation}\\
{Kivel N.A. }\\
{\small \it Theoretical Department, St.-Petersburg Nuclear Physics
Institute,}\\
{\small \it Gatchina, St.-Petersburg, 188350, Russian Federation}\\
{Stepanenko A.S.
\thanks{Permanent address: Theoretical Department, St.-Petersburg Nuclear
Physics Institute, Gatchina, St.-Petersburg, 188350, Russian Federation}}\\
{\small \it Service de Physique Th\'{e}orique, CE-Saclay, F-91191
Gif-sur-Yvette Cedex, France}\\
{Vasil'ev A.N.}\\
{\small \it Department of Theoretical Physics, St.-Petersburg University,}\\
{\small \it Ul'yanovskaya 1, Staryi Petergof, 198904, St.-Petersburg,
Russian Federation}}
\date{  }

\begin{document}
\maketitle
\thispagestyle{headings}
\begin{abstract}
        A proof of critical conformal invariance of Green's functions for a
        quite wide class of models possessing critical scale invariance is
        given. A simple method for establishing critical conformal
        invariance of a composite operator, which has a certain critical
        dimension, is also presented. The method is illustrated with the
        example of the Gross--Neveu model and the exponents \et\ at order
        $1/n^3$, \Dl\ and $1/\nu$ at order $1/n^2$ are calculated with the
        conformal bootstrap method.
\end{abstract}
\def\thepage{\bf SPhT-93/016}
\def\rightmark{}
\newpage
\setcounter{page}{1}
\def\thepage{\arabic{page}}

\section{Introduction.}
        It is well known that the structure of renormalized field theory is
        determined by the renormalization group (RG) equations. One
        considerable problem then is the calculation of renormalization
        group functions.  Usually one can compute only a few orders in
        perturbation theory.  Hence, we do not have a complete knowledge of
        these functions.  Other methods, such as the large $N$ expansion,
        usually allow us to compute RG functions at leading order only
        because calculations become too complicated at higher orders.

        To overcome these difficulties other methods have been developed.
        One such approach \ci{SEP,Gr} uses critical scale invariance at the
        fixed point of renormalization group for calculation of critical
        exponents, which are values of RG functions at fixed point. With
        this method it is possible to derive the critical exponents at
        order $1/N^2$. Using then the exact $N$ dependence of RG functions
        at each order in perturbation theory one can reconstruct them at
        some high-loop approximation.

        In order to derive the critical exponents at order $1/N^3$ a more
        powerful technique, the conformal bootstrap method, can be applied
        (see, for example, \ci{Vas} and references given there). To use
        this method it is necessary that the model possesses conformal
        symmetry at the critical point. Up to now this method has been
        applied only to the calculation of the exponent $\eta$ at order
        $1/N^3$ in the non-linear $\sg$ model \ci{Vas} which has critical
        conformal symmetry.  The aim of this paper is a proof of critical
        conformal invariance for a quite wide class of models. An example
        will be provided by the Gross--Neveu model.

\section{Preliminaries and conventions.}
        The $U_N$--symmetric Gross--Neveu (GN) model describes a system of
        $N$ $d$--dimensional Dirac spinors $\psi,\psb\equiv \{\psi_a,
        \psb_a, \ a=1, \dots ,N\}$, its non-renormalized action being
\be
        S = \int\!{\rm d}x\ [\psb \plh \psi + g_0 (\psb \psi)^2/2]
\lab{1}
\ee
        (in what follows we consider the theory in Euclidean space and
        functional distribution is denoted by $\exp S$ without the minus in
        the exponent). In statistical physics the coupling constant $g_0$
        in \re{1} is a bare temperature. It is well known that the GN
        model has properties quite analogous to the non-linear \sg--model.
        It is renormalizable in two dimensions and its renormalized
        action in $d=2+\eps$ dimension is
\be
        S = \int\!{\rm d}x\ [Z_1\psb \plh \psi + gM^{-\eps}Z_2 (\psb
        \psi)^2/2] \ ,
\lab{2}
\ee
        where $g$ is a dimensionless renormalized coupling constant
        (renormalized temperature), $M$ is a renormalization mass and
        $Z_{1,2}$ are renormalization constants.

        If $g$ is small the fermions are massless due to a chiral symmetry.
        When $g$ increases the system has a second order phase transition
        at some critical point $g_*$ (fixed point of renormalization
        group). At this point the chiral symmetry is spontaneously broken
        and the fermions become massive. There is a critical scaling in the
        vicinity of $g_*$.  Critical dimensions $\Delta [F]$ of various
        quantities $F$ (fields and parameters) can be calculated in the
        frameworks of $2+\eps$, $4-\eps$ and \nn\ expansions with $n\equiv
        N\cdot\tr$, $\tr$ being a trace of the unit matrix in the space of
        $d$--dimensional spinors.  This trace does not have (and does not
        require) an exact expression through $d$ (often one takes $\tr
        =2,2,4$ for the integer dimensions $d=2,3,4$).

        The $2+\eps$ expansion can be derived from the perturbation
        calculation of RG functions \ci{Wet}. To generate a systematic \nn\
        expansion a scalar field \sg\ is introduced and the GN action
        \re{1} is written in the following form:
\be
        S = \int\!{\rm d}x\ [\psb\plh\psi-\sg^2/2g_0+\psb\psi\sg] \  .
\lab{3}
\ee
        The $4-\eps$ expansion can be derived from the model suggested in
        \ci{Zinn}.  This model differs from \re{3} by adding of terms like
        $(\pl\sg)^2$ and $\sg^4$. In the framework of the \nn\ expansion
        such terms play no role in the IR limit. Hence, the models
        \ci{Zinn} and \re{3} have the same \nn\ expansion. But the model
        \ci{Zinn} is renormalizable in $d=4$ and it allows us to obtain the
        $4-\eps$ expansion of the same critical exponents with the standard
        RG technique.

        In what follows we shall consider \nn\ expansions in arbitrary
        dimension of space $d$ and the following notations will be used:
\be
        N\tr \equiv n\ , \ \ d \equiv 2\mu\ , \ \ B_z\equiv \psi(z) +
        \psi(\mu-z)\ ,
\lab{4}
\ee
        where $\psi(z)=\Gamma^\prime(z)/\Gamma(z)$ is the logarithmic
	derivative of the gamma--function $\Gamma(z)$. In the framework of
        the \nn\ expansion for the model \re{3}, the canonical dimensions
        of the fields $\vph \equiv \psi,\sg$ and temperature $\tau\equiv
        g_* - g$ are
\be
	d_\psi=d_{\psb}=\mu-1/2\ ,\ \ d_\sg =1\ ,\ \ d_\tau = 2(\mu-1)\ .
\lab{5}
\ee
        The notation $d[F]\equiv d_F$ means canonical dimension of quantity
        $F$. Let us parametrize the critical dimensions $\Dl_F=d_F+\gm_F^*$
        (canonical $+$ anomalous) as follows:
\be
	\Dps=\dps+\eta/2\ ,\ \ \Dsg=d_\sg-\eta-2\Dl\ ,\ \ \Dl_\tau \equiv
	1/\nu \equiv 2\ld{} \ \ (\ld{0}=\mu-1)\ ,
\lab{6}
\ee
        where $\eta$ and $\nu$ are the usual notations for the critical
        exponents, the quantity $2\Dl$ is a critical dimension of the
        vertex $\psb\psi\sg$ in the model \re{3}. For this model a
        connection between the quantity $\Dl_\tau$ and the critical
        dimension $\Dl[\sg^2]$ of a composite operator $F=\sg^2$ is
        expressed by the following relation
\be
	\Dl_\tau \equiv 1/\nu \equiv 2\ld{} = 2\mu-\Dl[\sg^2] \ .
\lab{7}
\ee
        Note that in the framework of the $2+\eps$ scheme for the purely
        fermion model \re{2} (without the field \sg) the quantity \Dsg\ can
        be calculated directly from the relation $\Dsg=\Dl [\psb\psi]= 2\mu
        -\Dl_m$ through the dimension of the composite operator
        $F=\psb\psi$ or through the dimension of the fermion mass $m$ for a
        massive generalization of the model \re{2} (this mass is generated
        by the substitution $\plh \rightarrow\plh+m_0$ with $m_0=mZ_m$).

        For the \nn\ expansion of any exponent $z$ its coefficients in
        powers of \nn\ are denoted by $z_k$:  $z=z_0+z_1/n+z_2/n^2+\dots$. In
        what follows the results for the three independent exponents $\eta,
        \Dl, \ld{}$ in \re{6} are presented.

\section{Main results for the Gross--Neveu model.}
        The first coefficients $\eta_1,\Dl_1,\ld{1}$ for the special case of
        dimension $d=2\mu=3$ have been calculated in \ci{Gat}; in our
        notations these results have the form:
\be
        \eta_1=8/3\pi^2\ ,\ \ \Dl_1/\eta_1=3/2\ ,\ \ \ld{1}/\eta_1=-2\ .
\lab{8}
\ee
	The same coefficients and $\eta_2$ for arbitrary dimension $d=2\mu$
        have been calculated in \ci{Gr}. Their results are
\be
	\left.
	\begin{array}{l}
        \eta_1=-{\displaystyle\frac{2\Gm(2\mu-1)}{\Gm(\mu)\Gm(\mu+1)
        \Gm(\mu-1)\Gm(1-\mu)}}\ ,\\[10pt]
        {\Dl}_1/\eta_1={\displaystyle\frac{\mu}{2(\mu-1)}}\ ,\ \
        {\ld{1}}/\eta_1=-(2\mu-1)\ ,\\[10pt]
        \eta_2/\eta_1^2=(B_{1-\mu}-B_1){\displaystyle\frac{(2\mu-1)}{
        (\mu-1)}}+{\displaystyle\frac{1}{2\mu}}-
        {\displaystyle\frac{\mu}{2(\mu-1)^2}}\ .
	\end{array}
	\right\}
\lab{9}
\ee
	We calculated the following coefficients $\Dl_2,\ld{2}$ and
	$\eta_3$:
\eqalph
\be
	\left.
        \begin{array}{c}
        {\Dl}_2/\eta_1^2 = {\displaystyle\frac{\mu}{2(\mu-1)}}\biggl[
        (B_{1-\mu}-B_1){\displaystyle\frac{(2\mu-1)}{(\mu-1)}}-3\mu
        B^\prime_1-\\[10pt]
        -2\mu-1-{\displaystyle\frac{6}{(\mu-1)}}-
        {\displaystyle\frac{4}{(\mu-
        1)^2}}\biggr] \ ,
        \end{array}
	\right\}
\lab{10a}
\ee
\be
	\left.
        \begin{array}{c}
        {\ld{2}}/\eta_1^2= {\displaystyle\frac{\mu}{2(\mu-1)}}\biggl[
        {\displaystyle\frac{8}{\eta_1(2-\mu)^2}}+
        [B^\prime_{2-\mu}-(B_{2-\mu}-B_1)^2]
        {\displaystyle\frac{4\mu(3-2\mu)}{(2-\mu)}} +\\[10pt]
        +(B_{2-\mu}-B_1)\left(-8\mu^2+2-{\displaystyle\frac{8}{(\mu-1)}}
        -{\displaystyle\frac{10}{(2-\mu)}}+
        {\displaystyle\frac{4}{(2-\mu)^2}}-{\displaystyle\frac{2}{\mu}}
        \right)+\\[10pt]
        +\mu B^\prime_1\left(6\mu-35+{\displaystyle\frac{22}{(2-\mu)}}
        \right) +8\mu^2-16\mu-2+\\[10pt]
        +{\displaystyle\frac{14}{(\mu-1)}}-{\displaystyle\frac{5}{
        (\mu-1)^2}}+{\displaystyle\frac{42}{(2-\mu)}}-
        {\displaystyle\frac{4}{(2-\mu)^2}}+
        {\displaystyle\frac{4}{\mu}}-{\displaystyle\frac{1}{\mu^2}}
        \biggl]
        \end{array}
	\right\}
\lab{10b}
\ee

\be
	\left.
        \begin{array}{c}
        \eta_3/\eta_1^3={\displaystyle\frac{1}{4}}\biggl[
        (B^\prime_1-B^\prime_{1-\mu})
        {\displaystyle\frac{2(2\mu-1)^2}{(\mu-1)^2}}
        -I(\mu)B^\prime_1{\displaystyle\frac{6\mu^2}{(\mu-1)}}+ \\[10pt]
        + B^\prime_1\left(8-\mu+{\displaystyle\frac{16}{(\mu-1)}}+
        {\displaystyle\frac{9}{(\mu-1)^2}}\right) - \\[10pt]
        -B^\prime_1(B_{1-\mu}-B_1){\displaystyle\frac{12\mu^2}{(\mu-1)}} +
        (B_{1-\mu}-B_1)^2{\displaystyle\frac{6(2\mu-1)^2}{(\mu-1)^2}}+
        \\[10pt]
        +(B_{1-\mu}-B_1)\left(-4\mu-6+{\displaystyle\frac{6}{\mu}}
        -{\displaystyle\frac{28}{(\mu-1)}}-
        {\displaystyle\frac{40}{(\mu-1)^2}} -
        {\displaystyle\frac{14}{(\mu-1)^3}}\right)+\\[10pt]
        -4\mu-6+{\displaystyle\frac{2}{\mu^2}}-
        +{\displaystyle\frac{9}{(\mu-1)}}+
        {\displaystyle\frac{3}{(\mu-1)^2}}+
        {\displaystyle\frac{12}{(\mu-1)^3}}
        +{\displaystyle\frac{5}{(\mu-1)^4}}\biggl]\ ,
        \end{array}
	\right\}
\lab{10c}
\ee
\eqarab
        where the notations \re{4} have been used and $I(\mu)$ is a
        function which cannot be expressed explicitly through $\Gm(z)$ and
        its derivatives. A definition of this function will be given below.
        The expression for $\eta_3/\eta_1^3$ in the non-linear \sg--model
        \ci{Vas}, which is analogous to \re{10c}, also contains this
        function. We write below this expression for the \sg--model in
        the notations \re{10c}, correcting a misprint in \ci{Vas}:
\be
	\left.
        \begin{array}{c}
        {\et_3}/\et_1^3 =
        -B^\prime_1I(\mu){\displaystyle\frac{3\mu^2(\mu-1)
        (4\mu-5)}{2(2-\mu)^2}} +
        {\displaystyle\frac{2\mu^2(\mu-1)(2\mu-3)^2}{3(2-\mu)^3}}\times
        \\[10pt]
        \times\left[3(B_{2-\mu}-B_1)(B^\prime_{2-\mu}-B^\prime_1)-
        B^{\prime\prime}_{2-\mu}+ B^{\prime\prime}_1-(B_{2-\mu}-B_1)^3
        \right] +\\[10pt]
        +{\displaystyle\frac{1}{2}} \biggl\{8\mu^2+26\mu+70 -
        {\displaystyle\frac{177}{(2-\mu)}}+
        {\displaystyle\frac{67}{(2-\mu)^2}}+
        {\displaystyle\frac{58}{(2-\mu)^3}}-\\[10pt]
        -{\displaystyle\frac{16}{(2-\mu)^4}}+
        {\displaystyle\frac{9}{(\mu-1)}}+
        {\displaystyle\frac{1}{(\mu-1)^2}}+
        {\displaystyle\frac{1}{\mu^2}}+(B_{2-\mu}-B_2)\times\\[10pt]
        \times\left[4\mu^2+14\mu+66-{\displaystyle\frac{187}{(2-\mu)}}+
        {\displaystyle\frac{102}{(2-\mu)^2}}+
        {\displaystyle\frac{16}{(2-\mu)^3}}+
        {\displaystyle\frac{2}{(\mu-1)}}
        +{\displaystyle\frac{3}{\mu}}\right]+\\[10pt]
        +(B_{2-\mu}-B_2)^2\left[20-{\displaystyle\frac{50}{(2-\mu)}}+
        {\displaystyle\frac{32}{(2-\mu)^2}}\right]-\\[10pt]
        -B^\prime_1\left[7\mu^2-10\mu-45+
        {\displaystyle\frac{127}{(2-\mu)}}-
        {\displaystyle\frac{64}{(2-\mu)^2}}-
        {\displaystyle\frac{48}{(2-\mu)^3}}+
        {\displaystyle\frac{32}{(2-\mu)^4}}\right] +\\[10pt]
        +B^\prime_{2-\mu}\left[8\mu^2+8\mu+14-
        {\displaystyle\frac{30}{(2-\mu)}}
        \right]-\\[10pt]
        -(B_{2-\mu}-B_2)B^\prime_1\left[-2\mu^2-13\mu-45+
        {\displaystyle\frac{136}{(2-\mu)}}-
        {\displaystyle\frac{108}{(2-\mu)^2}}+
        {\displaystyle\frac{32}{(2-\mu)^3}}\right]
        \biggl\}\ ,
        \end{array}
	\right\}
\lab{11}
\ee
        in the notations \re{4}. In the denominator of the second term of
        the expression for $\eta_3$ in \ci{Vas} (eq.(22) of \ci{Vas})
        $(2-\mu)^2$ appeared instead of $(2-\mu)^3$. The value of $\eta_3$
        for the integer dimension $d=2\mu=3$ is not affected by the
        misprint but it is essential for comparing \eps--expansions.  The
        expression \re{11} is more complicated than \re{10c} and this is a
        general rule:  calculations in the GN model are simpler because the
        number of diagrams is smaller due to the non-existence of odd
        fermion cycles.

        The function $I(\mu)$ is expressed through the self-energy massless
        diagram showed in Fig.1 on which a line with index $a$ means a
        propagator like $|x-y|^{-2a}$. According to its dimension the
        diagram is a simple line with index $\mu-1+\Dl$ multiplied by the
        coefficient $\Pi(\mu,\Dl)$ which is the ``value'' of the diagram.
        The quantity $I(\mu)$ is defined from $\Pi(\mu,\Dl)$ by the
        relation $I(\mu)=d\ln\Pi(\mu,\Dl)/d\Dl|_{\Dl=0}$ (for $\Dl=0$ the
        diagram is calculated exactly:  $\Pi(\mu,0)=3\pi^{2\mu} B^\prime_{
        \mu-1} \Gm(\mu-1) \Gm(2-\mu)/\Gm(2\mu-2)$).  In \ci{Vas} the value
        of $I(\mu)$ for the dimension $d=2\mu=3$ and the first terms of
        \eps--expansions around $d=2$ and $d=4$ are given:  $I(3/2)=
        3\psi^{\prime\prime} (1/2)/2 \pi^2+2\ln 2$, $I(1+\eps)=-2/3\eps
        +O(\eps)$, $I(2-\eps)=O(\eps)$.

        The result \re{9} was obtained in \ci{Gr} with the technique of
        self-consistent equations for propagators suggested in \ci{SEP}.
        The result \re{11} at order $1/n^3$, which was the unique one until
        quite recently, was obtained in \ci{Vas} with the conformal
        bootstrap method. In this method not only propagators but 3-point
        vertices also are dressed and this essentially simplifies
        calculations at high orders in \nn.  We obtained the results (10)
        with the same method, first proving conformal invariance. Technical
        details of the calculations will be published elsewhere \ci{TMF}.

        Here we restrict ourself to a discussion of the results (10) and
        their consequences for RG--functions in the model \re{3} (section
        6).  Now we give a proof of the critical conformal invariance for a
        quite wide class of models which includes the GN model.

\section{General proof of critical conformal invariance.}
        Let us consider an arbitrary multiplicatively renormalizable local
        model. The simplest dimensional regularization and MS scheme are
        supposed. Let us also suppose that the model has a renormalization
        group fixed point at which a critical scaling occurs. Hence, the
        corresponding massless model has a critical scale symmetry. The
        questions which we ask ourself are 1) would the model under some
        assumptions have a conformal symmetry of Green's functions
        $\langle\vph\vph\dots\rangle$ for simple fields only and 2) would
        it have conformal symmetry of Green's function $\langle
        F\vph\vph\dots\rangle$ with one insertion of a some composite
        operator $F$ (we assume the conformal symmetry with the same
        critical dimensions as in the scale symmetry).

        Let $S(\vph)=\int\!{\rm d}x {\cal L}(x)$ be the renormalized action
        of the massless model, \vph\ being a set of fields (needed indices
        and summations over them are implied). Let us also assume that
        ${\cal L} = {\cal L}_0+\v$, where \LL{0}\ is the quadratic part of
        the lagrangian, \v\ the interaction containing higher powers of
        \vph.  The quantity
\be
        W(A)=\ln\int\!{\rm D}\vph\ \exp[S(\vph)+\vph A]
\lab{12}
\ee
        is a generating functional of renormalized connected Green's
        functions for fields \vph\ and the quantity
\be
        \laa F\raa\equiv{{\int\!{\rm D}\vph\ F\exp[S(\vph)+\vph A]} \over
        {\int\!{\rm D}\vph\ \exp[S(\vph)+\vph A]}}
\lab{13}
\ee
        with arbitrary composite operator $F\equiv F(\vph)$ (local or
        non-local) is a generating functional of connected Green's
        functions $\langle F\vph\vph\dots\rangle$ with one operator
        $F$ insertion and any number of fields \vph. The source $A(x)$ in
        \re{12} and \re{13} is a functional argument.

        Let us consider an arbitrary group of transformations on the fields
        $\vph(x)\rightarrow\vph^\prime(x)$ which has an infinitesimal form
        like:
\be
        \vph^\prime(x)=\vph(x)+\om_{\al{}}T^{\al{}}\vph(x)\ ,
\lab{14}
\ee
        where $\om_{\al{}}$ are infinitesimal parameters, $T^{\al{}}$ are
        the corresponding generators acting on the fields \vph, \al{}\ is a
        numbering index. Summation over repeated indices is implied. We are
        interested in the scale and the special conformal transformations
        for which we have
\eqalph
\be
        T^{\al}\equiv T=\delta_{\vph}+x\pl \qquad \mbox{(scale)}
\lab{15a}
\ee
\be
        T^{\al}\equiv T^k=x^2\pl_k-2x_k x_s \pl_s - 2\delta_\vph x_k + 2
        \Sigma_{ks} x_s \qquad \mbox{(conformal)}
\lab{15b}
\ee
\eqarab
        where $\delta_\vph$ is the free given dimension of $\vph$,
        $\Sigma_{ks}$ is the spin part of the rotation generator (it is a
        known matrix acting on the $\vph$ field indices). With each
        transformation group one can associate the corresponding Noether
        currents \J{\al}{i}: $J_i = x_k\th{k}{i} + \delta_\vph\Ph{i}\vph$
        for the scale transformations and $\J{k}{i}=(x^2\dl_{ks}- 2x_kx_s)
        \th{s}{i} - 2x_k\delta_{\vph} \Ph{i}\vph + 2x_s\Ph{i} \Sigma_{ks}
        \vph$ for the special conformal ones. Here we use the following
        notations $\Ph{i}\equiv \pl\LL{}/\pl(\pl_i\vph)$ and $\th{k}{i}
        \equiv \Ph{i} \pl_k\vph-\dl_{ik}\LL{}$ (the stress-energy tensor).
        Note that in these notations we have $\dl S(\vph) /\dl\vph=
        \pl\LL{} /\pl\vph-\pl_i\Ph{i}$. We do not distinguish between upper
        and lower indices, their position is for convenience only.

        The following Ward identity is true for any group of
        transformations even if the system does not have the corresponding
        symmetry
\be
        \pl_i\J{\al}{i}(x)+\bigl[ \dl S(\vph)/\dl\vph(x) \bigr] \cdot \T{\al}
        \vph(x)+\NN{\al}(x)=0\ ,
\lab{16}
\ee
        where \NN{\al}\ is a ``breaking'' operator. If the symmetry is
        exact then $\NN{\al}=0$. From \re{16} we have
\be
        \dl^\al S\equiv\int\!{\rm d}x\ \bigl[ \dl S(\vph)/\dl\vph(x) \bigr]
        \T{\al} \vph(x)=-\int\!{\rm d}x\ \NN{\al}(x)\ .
\lab{17}
\ee
        On other hand from the following Schwinger--Dyson equation
\be
        0=\int\!{\rm D}\vph\ \dl\bigl\{\T{\al} \vph(x) \cdot \exp[ S(\vph)+
        A\vph] \bigr\}/\dl\vph(x)\ ,
\lab{18}
\ee
        one can derive the following relation $\laa\dl^\al S+\int\!{\rm d}x
        A(x) T^\al\vph(x)\raa=0$ (here the notation $\dl^\al\equiv\int\!
        {\rm d}x\ \T{\al}\vph(x) \dl/\dl\vph(x)$ is used) integrating over
        $x$ and neglecting terms like $\dl(0)$ (due to the dimensional
        regularization in the massless theory).  From this relation and
        \re{17} substituting $\vph\rightarrow\dl/\dl A$ it follows
\be
        \DA{\al}W(A)=\int\!{\rm d}x\ \laa \NN{\al} \raa\ ,
\lab{19}
\ee
        where $\DA{\al}\equiv\int\!{\rm d}xA(x)T^\al[\dl/\dl A(x)]$. If
        $\NN{\al}=0$ equation \re{16} implies that the corresponding
        symmetry of Green's functions $\la \vph\vph\dots \ra$, but in the
        general case of non-zero \NN{\al}\ the symmetry is broken.  All the
        above relations are satisfied for any free chosen dimension
        $\delta$ in (15). The exact form of the breaking operators
        $\NN{\al}\equiv \{ N\ \mbox{(scale)},\ \NN{k}\ \mbox{(conformal)}
        \}$ are influenced by the choice of $\delta$. In what follows we
        assume that $\delta$ in (15) are the canonical dimensions
        $\delta_\vph=d_\vph$.  Then under some assumptions on \LL{}\ the
        following relation between $N$ and \NN{k} holds:
\be
        \NN{k}(x)=-2x_kN(x)\ .
\lab{20}
\ee
        The assumptions are the following:
\begin{description}
\item[1. ] \LL{}\ is local and invariant under translations and
        rotations.
\item[2. ] The quadratic part of \LL{0}\ is invariant under the scale and
        the conformal transformations with canonical dimensions $d_\vph$.
\item[3. ] The interaction \v\ contains explicitly the fields \vph\ but
        not their derivatives $\pl\vph$.
\end{description}

        With these assumptions, only the interaction \v\ contributes to
        \NN{\al}. It is then easy to find $\NN{\al}\equiv\{ N,N^k \}$
        exactly, looking for contributions of \v\ to the l.h.s.. We have
        $N=-\v d+ d_\vph [\pl\v/\pl\vph]\vph$ and $N^k=-2x_kN+2x_s[\pl\v/
        \pl\vph] \Sigma_{ks}\vph$. The latter contribution is a variation
        of \v\ under the purely spin rotations and it has to disappear with
        our assumptions (locality + rotation symmetry + absence of
        derivatives $\pl\vph$ in \v). Hence, the equation \re{20} is
        satisfied. For the scale operator we have $N\sim\v$ if \v\ consists
        of one term only and it is a linear combination of the terms in the
        general case.  This operator has the same symmetry as \LL\ and
        canonical dimension $d$.

        The relation \re{20} is a basis for further constructions. In order
        to analyze consequences of the relation \re{19} in the critical
        region the r.h.s. of \re{19} should be expressed in terms of
        renormalized composite operators $[F]_R$ because the
        non-renormalized ones have no meaning at $g=g_*$ due to the
        singularities of renormalization constants $Z(g)$ at $g=g_*$.
        For the scale operator $\NN{\al}\equiv N$ it is convenient to use
        as elements of the basis the following renormalized ones (which
        correspond to each independent component of the field \vph)
\be
        \vph(x)\dl S(\vph)/\dl\vph(x) = \left[ \vph(x)\dl S_{\rm bas}(\vph)
        /\dl\vph(x) \right]_R\ .
\lab{21}
\ee
        $S_{\rm bas}$ is a ``basis'' action which is given by the
        renormalized one $S$ after substitution $Z\rightarrow 1$, the
        notation $[M]_R$ for any monomial $M$ means the appropriate
        renormalized one. Note that just the r.h.s. gives a meaning to the
        operator in \re{21} at $g=g_*$ because the constants $Z$
        contained in the l.h.s. do not exist at $g=g_*$. The Schwinger
        equation which follows from \re{18} after the substitution $T^\al
        \rightarrow 1$ yields
\be
        \laa\vph(x)\dl S(\vph)/\dl\vph(x)\raa = -A(x)\dl W(A)/\dl A(x)\ ,
\lab{22}
\ee
        which proves UV-finiteness of the operator \re{21} and vanishing of
        its anomalous dimension. Note that the equation \re{21} is a
        consequence of the MS scheme.

        In general the expansion of the scale operator $N$ in terms of
        renormalized ones contains three kinds of contributions (we denote
        them by A, B, C respectively):
\be
        N=\sum_\vph a_\vph\vph\dl S/\dl\vph+\sum b[F]_R+\sum c\pl [F]_R
\lab{23}
\ee
        where $a,b,c$ are numeric coefficients depending on the coupling
        constant $g$. The first sum (A) contains all the operators of the
        type \re{21}, the other two are the remaining independent operators
        with (C) and without (B) external derivatives. There is one
        contribution of type \re{21} (A), one $[\vph^4]_R$ of type B and
        one $\pl^2[\vph^2]_R$ of type C in the simple $\vph^4$--model.
        After substitution of \re{23} into equation \re{19} for the scale
        operator, contributions of type C vanish due to the integration
        over $x$. Contributions of type A lead to a change of the field
        dimensions $d_\vph \rightarrow d_\vph+ a_\vph$ due to \re{22} and
        contributions of type B correspond to the breaking of scale
        invariance with the dimensions $d_\vph+a_\vph$.  Since we have
        assumed that the model has critical scale invariance with the
        dimensions $d_\vph+\gm_\vph^*$ then we have for the coefficients in
        \re{23}
\be
        a_\vph(g_*)=\gm_\vph^*\ ,\ \ b(g_*)=0\ .
\lab{24}
\ee
        The quantities $c(g_*)$ are not constrained.

        Let us consider now the equation \re{19} for the conformal case.
        The renormalization is not influenced by the multiplication by
        $-2x_k$. Hence, the expansion of the conformal operator $N^\al
        \equiv N^k$ in the r.h.s. of \re{19} in terms of the renormalized
        ones is given by \re{23} multiplied by $-2x_k$. Then from \re{15b}
        and \re{24} it follows that after the substitution into \re{19} the
        contributions of types A and B lead to the change $d_\vph
        \rightarrow \Dl_\vph$ of the canonical dimensions into the critical
        ones. Hence, in the absence of type C contributions the critical
        conformal invariance would be a consequence of the scale
        invariance.

        But the in general case type C contributions are present in \re{23}
        and can lead to a breaking of conformal symmetry. It is clear that
        after integration over $x$ in \re{19}, only the type C
        contributions containing only one external derivative like
        $\pl_k{\cal O}_k$ are dangerous because they are multiplied by one
        factor $x_k$.

        In the presence of such contributions the conformal relation
        \re{19} in the critical region yields
\be
        \DA{k}W(A)=c\int\!{\rm d}x\ \laa \OO{k}(x) \raa\ ,
\lab{25}
\ee
        where \OO{k}\ is a ``dangerous'' operator which does not depend
        explicitly on $x$, $c$ is a scalar numeric coefficient. Now in the
        definition (15) of generators \T{k}\ contained in \DA{k}, the
        dimensions $\delta_\vph$ should be taken as critical ones
        $\Dl_\vph$.

        The expression in l.h.s. of \re{25} is UV--finite and has identical
        canonical and critical dimensions (the functional $W$ is
        dimensionless and the operation \DA{k}\ decreases the dimension by
        the one). The coefficient $c$ is dimensionless at criticality
        because in the massless model it is proportional to some power of
        $\mu^\eps$ with the renormalization scale $\mu$. Hence, if $c\neq
        0$ the critical dimension $\Dl [\OO{k}]$ of the composite operator
        \OO{k}\ is obtained from \re{25} as follows:
\be
        \Dl[\OO{k}]=d-1\ .
\lab{26}
\ee

        Now let us consider an arbitrary renormalized composite operator
        $F(x)$, which is canonically conformal in the lowest order of the
        \eps\ expansion, with critical dimension $\Dl_F$. Then in analogous
        way we can derive the following relation
\be
        [\DA{k}+T_F^k]\laa F(x)\raa=c\laa \OO{k}(x) \raa\ ,
\lab{27}
\ee
        where $T_F^k$ is a generator like \re{15b} with a suitable
        dimension and spin rotation generator, \OO{k}\ is an analogue to
        \re{25} ``breaking'' operator which does not explicitly depend on
        $x$. If $F$ is non-scalar and is characterized by some indices then
        \OO{k}\ has the same. If $c\neq 0$ an analogue of \re{26} is now
        given by the following relation
\be
        \Dl[\OO{k}]=\Dl [F]-1\ .
\lab{28}
\ee

        Note that the canonical conformal invariance of the
        operator at $\eps=0$ is an essential restriction. For example, for
        the $\vph^4$--model with $\eps=4-d$, scalar operators containing
        only \vph\ and $\pl^2\vph$ are canonically conformal while
        operators like $\pl^4\vph,\pl^2\vph^2$ are not.

        A practical use of the formulae \re{25}--\re{28} is the following:
        possible candidates to be \OO{k}\ are defined by the symmetry and
        the canonical dimension $d_F^*\equiv d_F|_{\eps=0}= \Dl_F|_{\eps=0}
        $. If from the corresponding renormalized monomials one can
        construct a linear combination which satisfy the criterion \re{26}
        or \re{28} then the combination can be \OO{k}, i.e. the conformal
        symmetry can be broken (however the final answer is given by an
        exact calculation of the coefficient $c$ in the \eps\ expansion).
        But if there are no renormalized operators with the required
        dimension $\Dl[\OO{k}]$, then the r.h.s. of the corresponding
        equations \re{25} or \re{27} have to vanish due to $c(g_*)=0$. In
        this case the critical conformal symmetry is a consequence of scale
        invariance.

\section{Examples of models possessing critical conformal invariance.}
        Below we present some models for which the conformal symmetry of
        the Green's function of the simple fields is guaranteed (all the
        models satisfy the assumptions).

\noindent
{\large\bf 1.}
        For models with one scalar field \vph\ the operator \OO{k}\ in
        \re{25} is necessarily the gradient of another scalar $\pl_k F$.
        Such operator does not contribute after integration over $x$ in
        \re{25}.

\noindent
{\large\bf 2.}
        Let us consider the complex $\vph^4$--model with the fields $\vph,
        \vph^+$ and the interaction $\v\sim (\vph^+\vph)^2$. Gauge current
        $\J{\rm cal}{k}=[\vph^+\pl_k\vph-\pl_k\vph^+\cdot\vph]_R$ only is a
        candidate to be \OO{k}\ in \re{25}. It has no anomalous dimension
        \ci{Col}, is conserved, and is multiplicatively renormalizable
        (nothing can mix with it). Hence, \J{\rm cal}{k}\ satisfies the
        criterion \re{26}. But the dangerous term $\pl_k\J{\rm cal}{k}$ in
        r.h.s. of \re{23} is forbidden by charge symmetry, because it is
        odd under the charge transformation $\vph\leftrightarrow\vph^+$ and
        it cannot mix with the operator $N\sim\v\sim(\vph^+\vph)^2$ which
        is charge--even.  Note that the quantity $\pl_k\J{\rm cal}{k}$ is a
        difference of two operators like \re{21} for \vph\ and $\vph^+$.
        Its presence in r.h.s. of \re{23} would imply the inequality
        $\gm^*[\vph]\neq \gm^*[\vph^+]$ due to \re{24}, while conformal
        symmetry would be conserved in a wide sense. But the inequality
        $\gm^*[\vph]\neq \gm^*[\vph^+]$ is forbidden by the charge symmetry
        of the model.

\noindent
{\large\bf 3.}
        The same remark is valid for $O(n)$--symmetrical models with a
        $n$--component field $\vph\equiv\{\vph_a\}$, for instance the
        $O(n)$--$\vph^4$--model.

\noindent
{\large\bf 4.}
        The charge symmetry arguments are valid for other models possessing
        it, in particular, for models like the Gross--Neveu model \re{2}.
        For this model the basis of the renormalized monomials in r.h.s. of
        \re{23} consists of three operators $[(\psb\psi)^2]_R$, $[\psb \plh
        \psi]_R$ and $[\pl\psb\cdot\gm\psi]_R$.  Only the gauge current
        $\J{\rm cal}{k}= [\psb\gm_k\psi]_R$ gives use to dangerous
        $\pl_k\J{\rm cal}{k}$.  The contribution of $\pl_k\J{\rm cal}{k}$
        is prohibited by the parity of the breaking operator $N\sim\v \sim
        (\psb\psi)^2$ again. The inequality $\Dl[\psi]\neq \Dl[\psb]$ in
        the presence of conformal symmetry in the wide sense with unknown
        $\Dl[\psi]$ and $\Dl[\psb]$ is prohibited also by the exact form of
        the massless propagator $\la\psi\psb\ra\sim \widehat{x}/|x|^a$
        which is a conformal invariant only if $\Dl[\psi]=\Dl[\psb]$.  The
        last argument allows us to eliminate the dangerous operator
        $\pl_k\J{\rm cal}{k}$ without the reference to charge symmetry.
        This is important for calculations in arbitrary dimension $d$
        because the matrix of charge conjunction $C$, which is contained in
        the definition of the charge transformation $\{\psi, \psb\}
        \rightarrow \{\psi^\prime=C\psb, \psb^\prime=-\psi C^{-1}\}$ for
        the spinors, has no (similar to $\gm_5$) unique natural definition
        in arbitrary dimension $d$.

        It is clear that the above proof of the critical conformal symmetry
        of Green's functions for fields $\vph=\psb,\psi$ in the
        Gross--Neveu model \re{2} generalizes directly to other models with
        four--fermion interaction without derivatives.

\noindent
{\large\bf 5.}
        For an application of the conformal bootstrap method in \nn\
        calculations of critical dimensions it is necessary to generalize
        the proof of the critical conformal invariance to models like
        \re{3} with fields $\vph=\psi,\psb,\sg$. Moreover, we have to do
        that in \eps\ scheme, for which the standard theory of composite
        operator renormalization can be applied because we have used it in
        our proof. Note that the existence of conformal symmetry is an
        objective property and is not influenced by calculational schemes.
        The model \re{3} in the $2+\eps$ scheme is not suitable for our
        purpose because it is not multiplicatively renormalizable in the
        above scheme due to appearance of the second interaction
        $\sim(\psb\psi)^2$ as a counterterm.  Hence, the field \sg\
        apparently has no well defined critical dimension due to the mixing
        with the operator $\psb\psi$.  In the framework of the $2+\eps$
        renormalized model \re{3} the quantity \Dsg\ calculated in \nn\ is
        an eigenvalue of a $2\times 2$ matrix of the critical dimensions of
        the mixing operators \sg\ and $\psb\psi$ which complicates the
        analysis. It is simpler to consider a two--coupling generalization
        of the model \re{3} adding the second independent interaction
        $\sim(\psb\psi)^2$ which does not change the theory in the fermion
        sector. The two-coupling model is multiplicatively renormalizable
        and, hence, the field \sg\ has now a well defined dimension \Dsg.
        Just this dimension is calculated in \nn\ expansion. The proof of
        the conformal invariance of Green's functions for the fields
        $\sg,\psi,\psb$ is valid for the two-coupling model because the
        quadratic part $\sim\sg^2$ of the field \sg\ in the action is
        canonically conformal and the interactions do not contain the
        derivatives. It follows that the divergence of the gauge current is
        still an unique candidate to be a dangerous operator in r.h.s.  of
        \re{23}. But it is prohibited as we have seen.

        In order to calculate the exponent $1/\nu$ in \re{7} with the
        conformal bootstrap method we need to prove also the conformal
        invariance of some scalar composite operator $F$ with critical
        dimension $\Dl[F]=d-\Dl_\tau =d-1/\nu$. In the framework of $1/n$
        expansion for model \re{3} just operator $[\sg^2]_R$ has dimension
        $d-1/\nu$, but in the framework of $2+\eps$ scheme for the
        two-coupling generalization of model \re{3} some linear combination
        of four renormalized composite operators
        $\sg^2,\psb\psi\sg,(\psb\psi)^2,\psb\plh\psi-\pl\psb\cdot \gm\psi$
        does. As anomalous dimension $\gm_\tau^*$ is non-trivial then from
        \re{27}, \re{28} it follows that the operator $F$ is conformal one
        in $2+\eps$ scheme.  Indeed, from the analysis of canonical
        dimensions it follows that the gauge current \J{\rm cal}{k}\ only
        is a candidate to be the operator \OO{k}\ in \re{27}.  But it has
        no anomalous dimension, hence, it does not satisfy the criterion
        \re{28}. Because the conformal property is an objective one, hence,
        $[\sg^2]_R$ is a conformal operator in the framework $1/n$
        expansion.  In the same way it can be shown that the operator
        $[\vph^2]_R$ in the massless $O(n)$--$\vph^4$--model ($4-\eps$
        scheme) is critically conformal.  For the same model two of the
        three operators with $d^*_F=d_F|_{\eps=0}=4$ are conformal, but the
        third one ($\pl^2[\vph^2]_R$) is not.

        The scheme presented above gives us a quite simple way to analyze
        the problem of critical conformal invariance in any concrete model.
        One significant conclusion is that the fermion fields are not
        dangerous for conformal symmetry because the form of a massless
        spinor propagator, as for scalars but contrary to vectors, is
        solely defined by the standard symmetries without the conformal
        one. Note that vector fields are not dangerous by themselves (for
        instance, in the $U_N$--symmetrical Thirring model analogous to
        \re{3} the term $\sg^2\equiv\sg_i\sg_i$ is canonically conformal).
        However the presence of vector fields is usually correlated with the
        existence of a local gauge symmetry. In models like massless QED in
        an arbitrary gauge even the critical scale symmetry is broken due
        to the presence of the gauge parameter. In the transverse gauge
        $\pl_iA_i=0$ the critical scale symmetry occurs, but holds not the
        conformal symmetry because the constraint $\pl_iA_i=0$ is not
        preserved by conformal transformations. Note that we consider usual
        conformal transformations, and various attempts to generalize them
        \ci{Pal} are not included here because there are no real
        calculation applications of such methods.

\section{Discussions.}
        The expressions (10) can be tested by independently calculating
        $2+\eps$ and $4-\eps$ expansions of the corresponding quantities.
        In the $2+\eps$ scheme the quantities \re{6} are calculated from
        RG--functions of the model \re{3} with the following relations
\be
\begin{array}{c}
        {\Dps} = \dps +\gps^*\ ,\ \ \Dsg=\Dl [\psb\psi]=d[\psb\psi] + \gm^*
        [\psb\psi]\ ,\\
        {\Dl}_\tau=1/\nu=-\bt^\prime(g_*)\ \ (\dps=\mu-1/2\
        ,\ \ d[\psb\psi]=2\mu-1)\ ,
\end{array}
\lab{29}
\ee
        where $d_F\equiv d[F]$ are canonical dimensions of quantities $F$
        in the model \re{3} and $\gm^*$ are values of corresponding RG
        function $\gm(g)$ in the fixed point $g_*\sim\eps$. Three loop
        calculation of the RG functions in the MS scheme in the model
        \re{3} for $d=2+2\eps$ gives \ci{9}:
\eqalph
\be
        {\bt}=2\eps u-2u^2(n-2)+4u^3(n-2)+2u^4(n-2)(n-7)+O(u^5)
\lab{30a}
\ee
\be
        {\gps}=u^2(n-1)-u^3(n-1)(n-2)+O(u^4)
\lab{30b}
\ee
\be
        {\gm} [\psb\psi]=-2u(n-1)+2u^2(n-1)+2u^3(n-1)(2n-3)+O(u^4)
\lab{30c}
\ee
\eqarab
        where $u\equiv g/4\pi$ is a more convenient variable.  From (30) it
        follows
\eqalph
\be
        {\Dps}=1/2+\eps+\eps^2{\displaystyle \frac{(n-1)}{(n-2)^2}}+
        \eps^3{\displaystyle\frac{(n-1)(6-n)}{(n-2)^3}}+O(\eps^4)
\lab{31a}
\ee
\be
        {\Dsg}\equiv\Dl[\psb\psi]=1+2\eps-2\eps{\displaystyle
        \frac{(n-1)}{(n-2)}}-2\eps^2{\displaystyle \frac{(n-1)}{(n-2)^2}}+
        2\eps^3{\displaystyle\frac{n(n-1)}{(n-2)^3}}+O(\eps^4)
\lab{31b}
\ee
\be
        {\Dl_\tau}\equiv 1/\nu=2\eps-{\displaystyle\frac{4\eps^2}{(n-2)}}
        -4\eps^3{\displaystyle\frac{(n-3)}{(n-2)^2}}+O(\eps^4)
\lab{31c}
\ee
\eqarab
        Expanding (31) in \nn\ and (10) in \eps\ we can see that both
        results are in agreement.

        $4-\eps$ expansions of the quantities \re{6} also correlates with
        RG functions of the model \ci{Zinn}, but in $\Dl_F=d_F +\gm_F^*$
        the canonical dimensions of the latter model are implied:
        $\dps=(d-1)/2$, $d_\sg=(d-2)/2$, $d_\tau=2$ (the parameter
        $\tau\equiv g-g_*$ is the coefficient of $\sg^2$ in the action
        \ci{Zinn}). A one loop calculation \ci{Zinn} gives ($d=4-2\eps$):
\be
\left.
\begin{array}{c}
        {\gps^*}={\displaystyle\frac{\eps}{n+6}}+O(\eps^2)\ ,\ \
        {\gm_\sg^*}={\displaystyle\frac{\eps n}{n+6}}+O(\eps^2)\ ,\\[15pt]
        {\Dl_\tau}\equiv1/\nu=2+\gm_\tau^*=2-{\displaystyle
        \frac{\eps}{3(n+6)}}\bigl[5n+6+\sqrt{n^2+132n+36}\,\bigr]+
        O(\eps^2)\ .
\end{array}
\right\}
\lab{32}
\ee
        Expanding \re{32} in \nn\ and (10) in $\eps=(4-d)/2$ we can see
        again the agreement between these results.

        The quantities (10) can be expanded to any order in \nn, what gives
        us information about further terms in the expansions of \bt\ and
        \gps\ functions. Using the method of \ci{Gr} we could find all
        five--loop contribution to $\gm_\psi(g)$, if we would know the
        \eps\ expansion of the function $I(1+\eps)$ at order $\eps^2$.

        While preparing this manuscript we received the preprint \ci{10}
        which contains the result for $\Dl_2$ that is in agreement with
        \re{10a}.

\vspace{1cm}

        One of the authors thanks Dr.~R.~Lacaze for providing ref.\ci{10}.

\newpage


\unitlength=2.00mm
\special{em:linewidth 0.6pt}
\linethickness{1.0pt}
\begin{center}
\begin{picture}(70.00,20.00)(0,0)
        \put(1.00,10.00){\circle*{1.00}}
        \put(21.00,10.00){\circle*{1.00}}
        \emline{1.00}{10.00}{1}{11.00}{17.00}{2}
        \emline{1.00}{10.00}{3}{11.00}{3.00}{4}
        \emline{11.00}{17.00}{5}{21.00}{10.00}{6}
        \emline{11.00}{3.00}{7}{21.00}{10.00}{8}
        \emline{11.00}{3.00}{9}{11.00}{17.00}{10}
        \put(6.00,15.50){\makebox(0,0)[cc]{${}_1$}}
        \put(6.00,4.50){\makebox(0,0)[cc]{${}_1$}}
        \put(17.00,4.00){\makebox(0,0)[cc]{${}_{\mu-1}$}}
        \put(17.00,16.00){\makebox(0,0)[cc]{${}_{\mu-1}$}}
        \put(15.00,10.00){\makebox(0,0)[cc]{${}_{\mu-1+\Dl}$}}
        \put(24.00,10.00){\makebox(0,0)[cc]{$\equiv$}}
        \put(30.00,10.00){\makebox(0,0)[cc]{$\Pi(\mu,\Dl)$}}
        \put(34.50,10.00){\makebox(0,0)[cc]{$\cdot$}}
        \put(36.00,10.00){\circle*{1.00}}
        \put(46.00,10.00){\circle*{1.00}}
        \emline{36.00}{10.00}{11}{46.00}{10.00}{12}
        \put(41.00,8.00){\makebox(0,0)[cc]{${}_{\mu-1+\Dl}$}}
        \put(48.00,9.50){\makebox(0,0)[cb]{,}}
        \put(62.00,10.00){\makebox(0,0)[cc]{$I(\mu)\equiv\left.
        {\displaystyle {d\over d\Dl} } \ln \Pi(\mu,\Dl) \right|_{\Dl=0}$}}
        \put(37.00,-1.00){\makebox(0,0)[cc]{\large Fig.1}}
\end{picture}
\end{center}

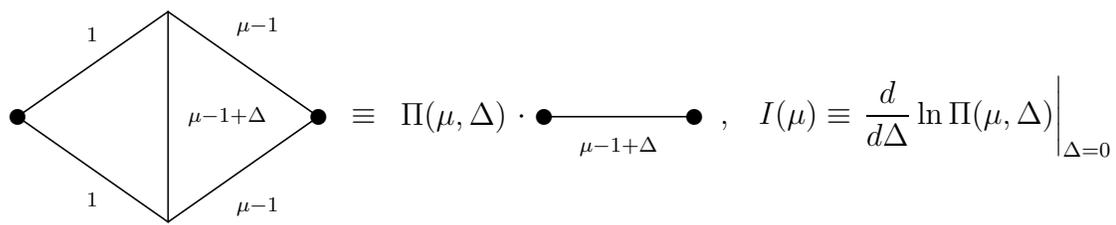
\end{document}